\PassOptionsToPackage{unicode}{hyperref}
\PassOptionsToPackage{hyphens}{url}
\documentclass[
]{article}
\usepackage{amsmath,amssymb}
\usepackage{lmodern}
\usepackage{iftex}
\ifPDFTeX
  \usepackage[T1]{fontenc}
  \usepackage[utf8]{inputenc}
  \usepackage{textcomp} 
\else 
  \usepackage{unicode-math}
  \defaultfontfeatures{Scale=MatchLowercase}
  \defaultfontfeatures[\rmfamily]{Ligatures=TeX,Scale=1}
\fi
\IfFileExists{upquote.sty}{\usepackage{upquote}}{}
\IfFileExists{microtype.sty}{
  \usepackage[]{microtype}
  \UseMicrotypeSet[protrusion]{basicmath} 
}{}
\makeatletter
\@ifundefined{KOMAClassName}{
  \IfFileExists{parskip.sty}{%
    \usepackage{parskip}
  }{
    \setlength{\parindent}{0pt}
    \setlength{\parskip}{6pt plus 2pt minus 1pt}}
}{
  \KOMAoptions{parskip=half}}
\makeatother
\usepackage{xcolor}
\usepackage{longtable,booktabs,array}
\usepackage{calc} 
\usepackage{etoolbox}
\makeatletter
\patchcmd\longtable{\par}{\if@noskipsec\mbox{}\fi\par}{}{}
\makeatother
\IfFileExists{footnotehyper.sty}{\usepackage{footnotehyper}}{\usepackage{footnote}}
\makesavenoteenv{longtable}
\ifLuaTeX
  \usepackage{selnolig}  
\fi
\IfFileExists{bookmark.sty}{\usepackage{bookmark}}{\usepackage{hyperref}}
\IfFileExists{xurl.sty}{\usepackage{xurl}}{} 
\hypersetup{
  hidelinks,
  pdfcreator={LaTeX via pandoc}}

\author{}
\date{}

\begin{document}

\textbf{Title: Practical Use of ChatGPT in Psychiatry for Treatment Plan
and Psychoeducation}

Authors: Farzan Vahedifard \footnote{Neuropsychiatry Research Team, Iran
  Sleep Disorders Clinic, Tehran, Iran. Email: farzanvahedi@yahoo.com},
Atieh Sadeghniiat Haghighi \footnote{Department of Psychiatry, Roozbeh
  Hospital, Tehran University of Medical Science, Tehran, Iran. Email:},
Tirth Dave \footnote{Internal Medicine, Bukovinian State Medical
  University, Chernivtsi, Ukraine. Email: tirth.snehal.dave@gmail.com} ,
Mohammad Tolouei \footnote{Neuropsychiatry Research Team, Iran Sleep
  Disorders Clinic, Tehran, Iran. Email: Tolouei@iransleep.com}, Fateme
Hoshyar Zare \footnote{Neuropsychiatry Research Team, Iran Sleep
  Disorders Clinic, Tehran, Iran. Email: Fhoshyar@iransleep.com}

Keywords: Artificial Intelligence; Psychiatry; Machine Learning; Deep
Learning; telepsychiatry; CBTi

\textbf{Abstract}

Artificial Intelligence (AI) has revolutionized various fields,
including medicine and mental health support. One promising application
is ChatGPT, an advanced conversational AI model that uses deep learning
techniques to provide human-like responses. This review paper explores
the potential impact of ChatGPT in psychiatry and its various
applications, highlighting its role in therapy and counseling
techniques, self-help and coping strategies, mindfulness and relaxation
techniques, screening and monitoring, education and information
dissemination, specialized support, group and family support, learning
and training, expressive and artistic therapies, telepsychiatry and
online support, and crisis management and prevention. While ChatGPT
offers personalized, accessible, and scalable support, it is essential
to emphasize that it should not replace the expertise and guidance of
qualified mental health professionals. Ethical considerations, such as
user privacy, data security, and human oversight, are also discussed. By
examining the potential and challenges, this paper sheds light on the
responsible integration of ChatGPT in psychiatric research and practice,
fostering improved mental health outcomes.

\textbf{Introduction}

Artificial Intelligence (AI) has recently emerged as a revolutionary
tool assisting doctors, researchers, coders, and many other
professionals. One particularly promising application of AI is in
medicine, where conversational AI transforms how mental health support
is provided, including psychiatry. One such model is ChatGPT, an
advanced language model that uses deep learning techniques to produce
human-like responses to natural language inputs (1). As natural language
processing and machine learning continue to advance, conversational AI
models like ChatGPT can potentially enhance psychiatric practice by
providing personalized, accessible, and scalable support through
interactive and empathetic conversations. Integrating conversational AI
in psychiatry is significant because it can provide on-demand support,
guidance, and resources to individuals seeking mental health assistance.
This creates a more user-friendly experience and breaks down barriers to
accessing mental health care (2).

ChatGPT's capabilities extend across various domains of psychiatry.
These include therapy and counseling techniques, self-help and coping
strategies, education and information dissemination, group and family
support, learning and training, mindfulness and relaxation techniques,
and crisis management and prevention. ChatGPT can provide therapeutic
conversations, offer coping mechanisms, and even simulate therapy
sessions (3). As we explore the applications of ChatGPT in psychiatry in
this paper, it is important to note that it serves as a valuable tool
within the therapeutic process, but should not replace the expertise and
guidance of qualified mental health professionals. ChatGPT can
complement the work of psychiatrists, psychologists, and counselors by
providing additional support, resources, and insights. However, it
should not be considered a replacement for the expertise and guidance of
qualified mental health professionals.

In this paper, we will explore the capabilities of ChatGPT and its
potential impact in various areas of psychiatry. We will also discuss
considerations and ethical implications associated with its use,
highlighting the importance of user privacy, data security, and human
oversight. By examining future research directions and challenges, we
aim to shed light on the exciting possibilities and responsible
integration of ChatGPT in psychiatric research and practice.

\textbf{Applications of ChatGPT in Psychiatry}

ChatGPT has numerous applications in psychiatry that can benefit both
patients and mental health professionals. These include therapeutic
conversations, mental health assessment, psychoeducation, supportive
chatbot, prevention and early intervention, continuity of care, crisis
intervention, decision support, and stigma reduction.

Regarding therapeutic conversations, ChatGPT can engage in interactive
and empathetic conversations, providing emotional support and helping
individuals explore their thoughts and feelings. ChatGPT can assist in
conducting initial assessments for mental health assessment by engaging
in structured interviews with patients. It can ask relevant questions
about symptoms, experiences, and emotions and summarize the collected
information to assist mental health professionals in making diagnoses
and treatment decisions.

ChatGPT can provide information about various mental health conditions,
treatment options, coping strategies, and self-help techniques through
psychoeducation as an educational tool. It can deliver personalized
materials based on user needs and questions. As a supportive chatbot,
ChatGPT can act as a virtual support companion for individuals with
mental health concerns. It can offer encouragement, reminders for
medication or therapy appointments, and suggestions for self-care
activities. Additionally, it can provide resources such as relaxation
techniques, mindfulness exercises, and guided breathing exercises.

Regarding prevention and early intervention, ChatGPT can be used in
mental health screening and early intervention programs. Engaging in
conversations with individuals, can identify early signs of distress or
risk factors for mental health problems. It can then provide appropriate
recommendations or refer users to mental health professionals for
further assessment. For continuity of care, ChatGPT can assist in
maintaining continuity by providing ongoing support between therapy
sessions. Patients can interact with the chatbot to reflect on their
progress, discuss challenges, and receive guidance on applying
therapeutic techniques learned during face-to-face sessions. During
mental health crises, ChatGPT can offer immediate support through crisis
intervention. It can engage in conversations to de-escalate distress,
provide coping strategies, and encourage individuals to seek appropriate
professional help when necessary. As a decision support tool, ChatGPT
can help mental health professionals make treatment decisions by
providing information on evidence-based practices, potential side
effects of medications and discussing treatment options based on
individual patient characteristics.

Finally, ChatGPT can play a role in reducing the stigma associated with
mental health through stigma reduction. It can engage in open and
non-judgmental conversations, creating a safe space for individuals to
discuss their mental health concerns. In the following sections, we will
explain the specific applications of ChatGPT in various domains of
psychiatry.

\begin{enumerate}
\def\labelenumi{\arabic{enumi}.}
\item
  \textbf{Therapy and Counseling Techniques}
\end{enumerate}

The field of mental health services has witnessed tremendous potential
in the application of AI, particularly in therapeutic and counseling
techniques. An area of great importance in this domain is the realm of
therapeutic conversations. AI models like ChatGPT offer users a platform
to seek emotional support during distress or anxiety. Interacting with
these models allows individuals to freely express their feelings and
concerns without fearing judgment, creating an environment that promotes
openness and self-reflection. This capability can be precious for those
who find it challenging to directly communicate their emotions to
another human due to stigma, anxiety, or other personal reasons.

Cognitive Behavioral Therapy (CBT) and Dialectical Behavioral Therapy
(DBT) are two widely used therapeutic approaches for addressing various
mental health conditions. While these therapies typically involve
sessions with trained professionals, AI can serve as a valuable source
of supplementary support. Users can rely on ChatGPT to assist them in
challenging negative thoughts or guiding them through distress tolerance
exercises. This capacity can be an important self-help tool, enabling
individuals to manage their symptoms between therapy sessions better.
Role-playing therapy, which involves practicing conversations or
anticipated confrontations that may be difficult, presents another area
where AI can offer benefits. Through role-playing scenarios with
ChatGPT, users can practice their responses and explore different
outcomes in a safe and pressure-free environment. Psychodynamic,
existential, and narrative therapies aim to help individuals understand
their past, find meaning and purpose in their lives, and reframe their
life stories in a positive manner, respectively.
ChatGPT\textquotesingle s ability to generate meaningful and
context-appropriate responses can aid users in exploring these
therapeutic themes effectively. Interpersonal Therapy (IPT) concentrates
on enhancing relationships and social interactions. While AI cannot
replace human interaction, it can support users in comprehending their
relational patterns and guide them for improving communication and
empathy.

It is crucial to emphasize that while ChatGPT and similar AI models can
provide significant value regarding emotional support and assistance in
therapeutic exercises, they should not substitute for professional
mental health services. AI can complement these services by providing
continuous support and tools for users to manage their mental health.
However, diagnosing and treating mental health conditions require
professional expertise and human intuition, which AI, in its present
stage, cannot replicate (4,5). Lastly, although privacy and data
security have significantly improved, users should be mindful of the
risks of sharing sensitive information. As AI evolves and advances, it
will likely play an increasingly prominent role in mental health
support, therapy, and counseling, extending its benefits to a larger
global population.

Table 1. Applications of ChatGPT in Therapy and Counseling Techniques

\begin{longtable}[]{@{}
  >{\raggedright\arraybackslash}p{(\columnwidth - 2\tabcolsep) * \real{0.5000}}
  >{\raggedright\arraybackslash}p{(\columnwidth - 2\tabcolsep) * \real{0.5000}}@{}}
\toprule()
\endhead
\textbf{Application} & \textbf{Example Command} \\
Therapeutic Conversations & "I\textquotesingle m feeling anxious. Can we
talk about it?"

Can you give me a strategy with the intense feelings that I have ? \\
Cognitive Behavioral Therapy Support & "Can you help me challenge this
negative thought I\textquotesingle m having?" \\
Dialectical Behavioral Therapy Support & "Can you help me with a
distress tolerance exercise?"

Is there a way to deal with daily stress and anxiety ? \\
Trauma-Informed Therapy Support & "Can you provide support for trauma
survivors?"

Can you help me with the frightening memories that I have

Can help me with the repetitive nightmaires about my accident \\
Role-Play Therapy & "Can we role-play a difficult conversation I need to
have with my boss?" \\
Existential Therapy Assistance & "Can you help me explore the meaning
and purpose in my life?"

"Can you explain the four main themes of existential therapy?"

"What is the concept of \textquotesingle authentic
self\textquotesingle{} in existential therapy?"

"what are the strategies for coping with the issue of death and
mortality in existatial therapy?"

"Could you help me understand \textquotesingle freedom and
responsibility\textquotesingle{} as understood in existential therapy?"

"What does \textquotesingle existential isolation\textquotesingle{} mean
and how can I cope with it?"

"Can you guide me through a self-reflection exercise based on
existential therapy principles?"

"How can I cultivate meaning in my life according to existential therapy
principles?" \\
Psychodynamic Therapy Assistance & "Can you help me explore how my past
might affect my current behavior?"

WHY DO I ALWAYS CHOOSE THE WRONG PARTNER? \\
Interpersonal Therapy Support &
\begin{minipage}[t]{\linewidth}\raggedright
"Can you help me understand how to improve my relationships?"

What are some strategies for setting bounderies and assertivitely
express my needs?\\
"What is Interpersonal Therapy about?"

"What happens during Interpersonal Therapy?"

"How does Interpersonal Therapy help when people fight or disagree?"

"How can Interpersonal Therapy help me when big changes happen in my
life?"

"How does Interpersonal Therapy help me if I\textquotesingle m feeling
sad about losing someone?"

"Can Interpersonal Therapy teach me better ways to talk and listen to
others?"

"What does a therapist do in Interpersonal Therapy?"

"Can you help me solve a problem like they do in Interpersonal Therapy?"

"How can Interpersonal Therapy help if I\textquotesingle m feeling very
sad or depressed?"\strut
\end{minipage} \\
Narrative Therapy Assistance & "how can I seprate myself from the
problem and see more objectively?"

"How can I rewrite my problem-saturated story in a more empowering way?"

"What is \textquotesingle re-authoring\textquotesingle{} in the context
of Narrative Therapy?"

"Can you guide me through a narrative-based self-reflection exercise?"

"What is \textquotesingle unique outcomes\textquotesingle{} in Narrative
Therapy and how can it be applied?" \\
\bottomrule()
\end{longtable}

\begin{enumerate}
\def\labelenumi{\arabic{enumi}.}
\setcounter{enumi}{1}
\item
  \textbf{Self-help and Coping Techniques}
\end{enumerate}

AI, including ChatGPT, has made remarkable progress in the realm of
self-help and coping techniques. This advancement has the potential to
transform how individuals access and engage in self-help strategies,
offering personalized, scalable, and available tools and techniques. One
such application is the development of personalized self-care plans.
Through conversations with ChatGPT, users can receive tailored
suggestions for stress management, encompassing relaxation techniques
and lifestyle adjustments that consider their preferences, lifestyle,
and specific stressors.

Another crucial aspect of self-help involves cultivating coping
mechanisms to manage emotions like anxiety, stress, and anger. Users can
seek suggestions from ChatGPT, exploring various strategies to
experiment with and identify what works best. This immediate and
personalized assistance can be precious when immediate access to a
therapist or counselor is impossible. Additionally, AI can support
behavioral activation, encouraging individuals to engage in enjoyable
activities to counter symptoms like depression. Users can combat low
moods and apathy by requesting activity suggestions aligned with their
interests and constraints.

AI also holds the potential to teach various therapeutic techniques,
such as somatic therapy, stress management, and anger management. Users
can participate in guided body awareness exercises or learn strategies
to manage their anger and stress. By providing these techniques, AI
equips users with tools to navigate their emotional states more
effectively. Furthermore, AI excels in emergency management, such as
during panic attacks, by offering immediate and calming assistance.
Users can ask ChatGPT for techniques to calm down during a panic attack,
providing on-the-spot support when it matters most.

Lastly, AI can contribute to self-esteem-building activities. By
suggesting positive affirmations, self-reflection exercises, and other
activities to boost self-esteem, AI becomes a valuable tool for personal
development and emotional well-being.

While AI can provide various resources and support for mental health
self-care, it is important to emphasize that it cannot replace the need
for professional mental health services.

Table 2. Applications of ChatGPT in Self-help and Coping Techniques

\begin{longtable}[]{@{}
  >{\raggedright\arraybackslash}p{(\columnwidth - 2\tabcolsep) * \real{0.3010}}
  >{\raggedright\arraybackslash}p{(\columnwidth - 2\tabcolsep) * \real{0.6990}}@{}}
\toprule()
\begin{minipage}[b]{\linewidth}\raggedright
Application
\end{minipage} & \begin{minipage}[b]{\linewidth}\raggedright
Example Command
\end{minipage} \\
\midrule()
\endhead
Personalized Self-Care Plans & "Can you create a personalized self-care
plan for managing my anxiety?" \\
& "What are some self-care activities I can incorporate into my daily
routine?" \\
& "How can I develop improve my mental well-being?" \\
& "Are there any self-care practices for reducing stress and promoting
relaxation?" \\
& "I need help creating a self-care plan to cope with my depression." \\
Developing Coping Mechanisms & "What are the correct coping mechanisms
for dealing with social anxiety?"

How coping mechanisms are mad

2- What are some ways to change our twisted coping mechanisms \\
& "What are some effective coping strategies for managing panic
attacks?" \\
& "I need coping techniques to overcome my fear of public speaking." \\
& "How can I develop healthy coping mechanisms to deal with stress?" \\
& "Can you suggest coping mechanisms for managing obsessive
thoughts?" \\
Encouraging Behavioral Activation & "What are some activities I can
engage in to boost my mood and motivation?" \\
& "Can you provide suggestions for behavioral activation techniques?" \\
& "I need ideas for activities that can help me overcome feelings of
lethargy." \\
& "How can I incorporate behavioral activation into my daily
routine?" \\
& "Can you suggest hobbies that promote behavioral activation and mental
well-being?" \\
Anger Management Techniques & "Can you teach me a technique for
controlling my anger when I feel provoked?"

What things that I can do with the sense of uncontrollable anger

What things that I can do with the sense of uncontrollable anger \\
& "What are some effective anger management strategies to avoid
aggression?" \\
& "I need help with controlling my aggression . Can you provide some
tips?" \\
& "How can I develop healthy ways to express and manage my anger?" \\
& "Can you suggest anger management exercises for calming down
quickly?" \\
Stress Management Techniques & "Can you teach me a stress reduction
technique that I can use daily?" \\
& "What are some effective stress management techniques for workplace
stress?" \\
& "How can I manage stress better in my personal life?" \\
& "I need techniques for stress management during exams. Can you
help?" \\
& "Can you suggest relaxation exercises to alleviate stress and
anxiety?" \\
Somatic Therapy Techniques & "Can you guide me through a grounding
theraputic exercise?" \\
& "What are some somatic therapy techniques for releasing tension?" \\
& "How can I practice body awareness for emotional healing?" \\
& "I need somatic exercise guidance for trauma recovery." \\
& "Can you suggest somatic techniques for managing chronic pain?" \\
Panic Attack Calming Techniques & "What can I do to during a panic
attack?" \\
& "Can you guide me through a breathing exercise for panic attack
relief?" \\
& "How can I prevent panic attacks and manage the symptoms?" \\
& "I need techniques to calm down quickly when experiencing a panic
attack." \\
& "Can you suggest grounding techniques for managing panic attack
symptoms?" \\
Self-Esteem Building Activities & "Can you suggest activities to improve
my self-esteem and self-worth?" \\
& "How can I build self-esteem through positive affirmations?" \\
& "What are some exercises for building confidence?" \\
& "I need activities that promote self-acceptance and self-love. Any
ideas?" \\
& "Can you provide tips for developing a positive self-image and
self-esteem?" \\
\bottomrule()
\end{longtable}

\begin{enumerate}
\def\labelenumi{\arabic{enumi}.}
\setcounter{enumi}{2}
\item
  \textbf{Mindfulness and Relaxation Techniques}
\end{enumerate}

ChatGPT and similar AI technologies are emerging as valuable tools for
mindfulness and relaxation techniques due to their accessibility,
customization potential, and scalability. They offer immediate and
personalized responses to individuals seeking to cultivate mindfulness,
manage sleep disorders, or learn relaxation techniques. Mindfulness, the
practice of focusing on present-moment experiences, has been recognized
for its mental health benefits, including stress reduction and
alleviating symptoms of anxiety and depression. ChatGPT can guide users
through short mindfulness exercises, serving as an accessible tool for
incorporating mindfulness into daily routines. These guided exercises
can be tailored to various lengths and complexity levels, providing
flexibility and personalization based on user preferences and
experience.

In addition to mindfulness exercises, ChatGPT can assist users with
biofeedback and relaxation techniques, such as deep breathing exercises.
These techniques effectively manage stress, anxiety, and other emotional
states. They can be easily integrated into a user\textquotesingle s
daily routine and utilized whenever needed, offering an on-demand
resource for promoting relaxation and well-being.

AI can also provide valuable assistance in sleep hygiene and sleep
disorder management. Sleep hygiene is crucial for maintaining physical
and mental health, yet many individuals struggle with sleep-related
issues. Users can seek tips from ChatGPT on improving sleep hygiene, and
receiving accessible and easy-to-understand advice for enhancing sleep
quality. For individuals experiencing sleep disorders like insomnia,
ChatGPT can offer guided sleep relaxation exercises to help them unwind
before bed and create a more conducive sleep environment. As AI
technology continues to advance, its role in mental health support is
expected to grow, potentially enhancing the accessibility and
personalization of care.

Table 3. Applications of ChatGPT in Mindfulness and Relaxation
Techniques

\begin{longtable}[]{@{}
  >{\raggedright\arraybackslash}p{(\columnwidth - 2\tabcolsep) * \real{0.5000}}
  >{\raggedright\arraybackslash}p{(\columnwidth - 2\tabcolsep) * \real{0.5000}}@{}}
\toprule()
\endhead
\textbf{Application} & \textbf{Example Command} \\
Mindfulness and Relaxation Techniques & What is mindfulness and how can
it help me ?

"Can you teach me a simple mindfulness exercise?"

"How can I relax when I\textquotesingle m feeling stressed?"

"Can you guide me through deep breathing?"

"How can I focus on the present moment?"

"Can you show me a way to relax my muscles?"

"What\textquotesingle s a good way to calm down quickly?"

"Can you help me learn to meditate?"

"How can mindfulness help me when I\textquotesingle m feeling anxious or
sad?"

"What can I do to sleep better?"

"Can you teach me a simple yoga pose for relaxation?" \\
Relaxation Techniques & Can you guide me through a deep breathing
exercise?" \\
Sleep Hygiene Education & Why Im having problem with my sleep recently ?

"What is sleep hygiene?"

How sleeping good affects my life ?

What are some first step tips for having a good sleep ?

"Why is good sleep important?"

"Tell me some habits for better sleep."

"How can I make my bedroom better for sleeping?"

"What foods and drinks can affect my sleep?"

"How can exercise help me sleep better?"

"What should I do if I can\textquotesingle t fall asleep?"

"How can a sleep schedule help me?"

"What\textquotesingle s the best way to wake up in the morning?"

"What is the role of naps in good sleep?" \\
Sleep Disorder Management & "what are the signs of sleep disorder?"

How can I know Im having problem with my sleep ? \\
\bottomrule()
\end{longtable}

\begin{enumerate}
\def\labelenumi{\arabic{enumi}.}
\setcounter{enumi}{3}
\item
  \textbf{Screening and Monitoring}
\end{enumerate}

Artificial Intelligence models, like ChatGPT, are increasingly being
leveraged in the field of psychiatry, where they hold considerable
potential to improve screening, monitoring, and overall patient care.
One significant benefit is their capacity for remote patient monitoring.
With ChatGPT, users can record their feelings and emotional states at
any time, creating a continuous log of emotional well-being. This
feature allows for more frequent check-ins and can help identify any
sudden changes or trends in a person\textquotesingle s mental health
status.

Regarding mental health screenings, ChatGPT can offer questions designed
to identify signs of various mental health conditions, such as
depression or anxiety. These questions are not diagnostic, but they can
help guide individuals toward seeking professional help if their
responses indicate potential mental health issues. Medication management
is another area where AI can provide significant support. ChatGPT can
help users keep track of their medication schedule, offering reminders
of when to take certain medications. This feature can benefit
individuals on multiple medications or those with conditions that may
impair memory.

In the diagnostic process, AI like ChatGPT can provide accessible,
user-friendly information on various psychiatric conditions, helping
individuals understand symptoms, prognosis, and treatment options.
However, it\textquotesingle s crucial to clarify that while ChatGPT can
provide information, it cannot make a clinical diagnosis. Kishimoto et
al. are conducting a study in which they are compiling an extensive
dataset of Japanese speech data. This dataset is meticulously labeled
with comprehensive information regarding psychiatric and neurocognitive
disorders. Their research aims to utilize natural language processing
techniques to quantify the linguistic characteristics associated with
these disorders. Ultimately, their goal is to develop objective and
user-friendly biomarkers that can aid in the diagnosis and assessment of
the severity of these conditions (6).

AI can also assist in assessing patient satisfaction and tracking
treatment progress. Users can report their satisfaction levels following
therapy sessions or other treatments, and over time, this feedback can
help healthcare professionals adjust their approach as needed. Moreover,
by tracking users\textquotesingle{} reported symptoms over time, AI can
help monitor the effectiveness of treatment and alert healthcare
professionals to any concerning trends. However, care must be taken to
ensure the ethical use of AI, particularly in terms of data security and
privacy, and users should always be made aware that AI, in its current
form, cannot replace the nuanced understanding and clinical judgment of
human healthcare providers.

Table 4. Applications of ChatGPT in Screening and Monitoring of Patients

\begin{longtable}[]{@{}
  >{\raggedright\arraybackslash}p{(\columnwidth - 0\tabcolsep) * \real{1.0000}}@{}}
\toprule()
\begin{minipage}[b]{\linewidth}\raggedright
\textbf{Mental Health Screening Questions}
\end{minipage} \\
\midrule()
\endhead
\textbf{"Have you noticed any changes in your sleep patterns
recently?"} \\
\textbf{"Do you often experience feelings of sadness or
hopelessness?"} \\
\textbf{"Have you lost interest or pleasure in activities you used to
enjoy?"} \\
\textbf{"Do you find it difficult to concentrate or make decisions?"} \\
\textbf{"Have you experienced changes in your appetite or weight?"} \\
\textbf{"Do you frequently feel restless or have trouble sitting
still?"} \\
\textbf{"Have you had recurring thoughts of death or suicide?"} \\
\textbf{"Do you feel tired or lacking in energy most of the time?"} \\
\textbf{"Have you been experiencing excessive worry or anxiety?"} \\
\textbf{"Do you often feel irritable or easily angered?"} \\
\bottomrule()
\end{longtable}

\begin{enumerate}
\def\labelenumi{\arabic{enumi}.}
\setcounter{enumi}{4}
\item
  \textbf{Education and Information}
\end{enumerate}

ChatGPT and similar AI technologies offer a wide range of opportunities
in the realm of education and information dissemination, especially in
the context of mental health. The ability to deliver immediate and
tailored responses makes them dynamic and flexible educational tools
that can cater to individual needs and inquiries.

One notable application is psychoeducation, where ChatGPT can provide
comprehensive information about various mental health conditions. Users
can ask specific questions about disorders like bipolar disorder and
receive detailed responses, thereby enhancing their understanding of
these conditions. Another valuable application is disseminating mental
health information, including the latest research. Users can inquire
about recent findings or advancements in treating mental health
conditions, and ChatGPT can provide summaries and explanations of the
latest research. This accessibility to up-to-date information empowers
users, enabling them to engage in their care and treatment decisions
actively.

AI can also play a significant role in reducing the stigma surrounding
mental health. ChatGPT contributes to dispelling myths and
misconceptions by providing accurate and non-judgmental information.
Users can seek advice on combating stigma in their communities,
promoting understanding and acceptance of mental health issues.

Furthermore, AI can assist in educating about the social determinants of
health. Understanding how factors like poverty impact mental health is
crucial for building more effective and equitable mental health support
systems. Users can ask ChatGPT to explain these relationships,
facilitating a more comprehensive understanding of mental health.

Lastly, cultural competency in mental health is a critical area where AI
can provide valuable insights. Culture influences the experience and
expression of mental health conditions and attitudes towards treatment.
ChatGPT can help users comprehend these cultural nuances, fostering a
more culturally sensitive approach to mental health.

Table 5. Applications of ChatGPT in Education and Information Sector

\begin{longtable}[]{@{}
  >{\raggedright\arraybackslash}p{(\columnwidth - 2\tabcolsep) * \real{0.3231}}
  >{\raggedright\arraybackslash}p{(\columnwidth - 2\tabcolsep) * \real{0.6769}}@{}}
\toprule()
\begin{minipage}[b]{\linewidth}\raggedright
Application
\end{minipage} & \begin{minipage}[b]{\linewidth}\raggedright
Example Command
\end{minipage} \\
\midrule()
\endhead
Psychoeducation & "Can you explain what bipolar disorder is?" \\
& "What are the symptoms of attention-deficit/hyperactivity disorder
(ADHD)?" \\
Disseminating Mental Health Information & "What is the latest research
on depression treatment?" \\
& "Are there any new breakthroughs in the treatment of anxiety
disorders?" \\
Stigma Reduction Education & "Can you provide me with information to
help reduce the stigma around mental illness in my community?" \\
& "Can you provide statistics on the prevalence of mental health
disorders to help combat misconceptions?" \\
Social Determinants of Health Education & "Can you explain how poverty
impacts mental health?" \\
& "How does access to healthcare services affect mental well-being?" \\
Cultural Competency in Mental Health & "Can you explain how my culture
might impact my mental health?" \\
& "Can you provide examples of how cultural norms can influence the
perception of mental illness?" \\
Test Preparation Assistance & "Can you suggest effective strategies for
improving my math problem-solving skills?" \\
Essay Writing Guidance & "How do I properly cite sources in an APA
format essay?" \\
Research Paper Assistance & "Can you help me narrow down my research
topic on environmental sustainability?" \\
Career Guidance & "What skills are in high demand in the field of data
science?" \\
College Application Support & "What extracurricular activities are
considered valuable for college applications in the field of
engineering?" \\
\bottomrule()
\end{longtable}

\begin{enumerate}
\def\labelenumi{\arabic{enumi}.}
\setcounter{enumi}{5}
\item
  \textbf{Specialized Support}
\end{enumerate}

ChatGPT\textquotesingle s role in psychiatry is expanding significantly,
encompassing specialized support for various demographics and
conditions. Its ability to offer tailored resources, techniques, and
strategies plays a crucial role in promoting positive mental health
outcomes for diverse populations. In the realm of substance abuse
recovery, ChatGPT can provide valuable coping strategies to manage
cravings, offering support to individuals on their path to recovery. For
instance, someone struggling with alcohol addiction can ask for
techniques to handle alcohol cravings, and ChatGPT can suggest methods
like distraction, mindfulness, or reaching out to a support network.

For individuals dealing with Post-Traumatic Stress Disorder (PTSD),
ChatGPT can provide assistance by explaining and guiding users through
grounding techniques to manage potential triggers. By seeking guidance
on grounding techniques, users can receive instructions from ChatGPT,
such as focusing on physical sensations or engaging in simple mental
exercises to anchor themselves in the present. Another area where
ChatGPT proves helpful is in supporting phobia desensitization. It can
guide users through imagined exposure to their phobias, gradually
assisting them in reducing their fear response. For example, if someone
fears spiders, ChatGPT can help them navigate the process of facing and
overcoming that fear.

In the case of conditions like eating disorders or body dysmorphic
disorder, ChatGPT can offer strategies to manage negative body image
thoughts or challenge distorted beliefs about one\textquotesingle s
appearance. These tools can complement professional treatment and
support users in their recovery journey.

ChatGPT can also provide assistance with managing Obsessive-Compulsive
Disorder (OCD) by suggesting techniques to handle obsessive thoughts. In
geriatric psychiatry, it can offer strategies to cope with issues like
loneliness in a nursing home environment. Furthermore,
ChatGPT\textquotesingle s ability to provide resources tailored to
specific groups, such as perinatal and postpartum mental health support,
LGBTQ+ mental health resources, veteran mental health support, and
refugee mental health resources, enhances access to specialized care for
these populations.

Grief and bereavement support, domestic violence support, youth mental
health support, and childhood trauma support are also within the
capabilities of ChatGPT. Whether guiding a user through a
grief-processing exercise or providing resources for domestic violence
survivors or individuals coping with childhood trauma, ChatGPT offers
valuable assistance.

Table 6. Applications of ChatGPT in Providing Specialized Support

\begin{longtable}[]{@{}
  >{\raggedright\arraybackslash}p{(\columnwidth - 2\tabcolsep) * \real{0.3687}}
  >{\raggedright\arraybackslash}p{(\columnwidth - 2\tabcolsep) * \real{0.6313}}@{}}
\toprule()
\begin{minipage}[b]{\linewidth}\raggedright
Application
\end{minipage} & \begin{minipage}[b]{\linewidth}\raggedright
Example Command
\end{minipage} \\
\midrule()
\endhead
Substance Abuse Recovery Support & "What are some coping strategies for
alcohol cravings?" \\
& "How can I resist cravings to use drugs at social events?" \\
PTSD Management Support & "Can you help me understand grounding
techniques for when I\textquotesingle m feeling triggered?" \\
& "What are some relaxation exercises to reduce anxiety during
flashbacks?" \\
Phobia Desensitization Support & "Can you guide me through an imagined
exposure to my phobia of spiders?" \\
& "What are some gradual steps I can take to overcome my fear of
heights?" \\
Eating Disorder Recovery Support & "What are some strategies to manage
negative body image thoughts?" \\
& "How can I develop a healthy relationship with food during
recovery?" \\
Body Dysmorphic Disorder Support & "Can you help me challenge my
negative beliefs about my appearance?" \\
& "What are some techniques to reduce excessive mirror checking
behaviors?" \\
Obsessive-Compulsive Disorder Management & "What are some techniques to
manage obsessive thoughts?" \\
& "How can I create a structured routine to minimize compulsive
behaviors?" \\
Geriatric Psychiatry Assistance & "Can you provide me with strategies to
cope with the loneliness I feel living in a nursing home?" \\
& "What are some activities I can engage in to improve my cognitive
function?" \\
Perinatal and Postpartum Mental Health Support & "Can you provide me
with resources for postpartum depression?" \\
& "What are some self-care strategies for new mothers?" \\
LGBTQ+ Mental Health Support & "What are some mental health resources
specifically for LGBTQ+ individuals?" \\
& "Can you help me navigate the coming out process to my family and
friends?" \\
Veteran Mental Health Support & "What are some strategies to cope with
the stress of transitioning to civilian life?" \\
& "Can you provide me with resources for veterans seeking mental health
support?" \\
Refugee Mental Health Support & "Can you provide me with mental health
resources for refugees in my area?" \\
& "What are some coping mechanisms for refugees experiencing trauma?" \\
Grief and Bereavement Support & "Can you guide me through a
grief-processing exercise?" \\
& "What are some strategies to cope with the loss of a loved one?" \\
Domestic Violence Support & "Can you provide me with resources for
domestic violence survivors?" \\
& "What are some safety planning tips for individuals experiencing
domestic violence?" \\
Youth Mental Health Support & "What are some ways to manage school
stress?" \\
& "How can I improve my self-esteem as a teenager?" \\
Childhood Trauma Support & "Can you provide me with resources to help
cope with childhood trauma?" \\
& "What are some grounding techniques for managing emotional
flashbacks?" \\
\bottomrule()
\end{longtable}

\begin{enumerate}
\def\labelenumi{\arabic{enumi}.}
\setcounter{enumi}{6}
\item
  \textbf{Group and Family Support}
\end{enumerate}

ChatGPT goes beyond individual support and extends its capabilities to
group and family settings, offering opportunities to improve
communication, facilitate therapy, and enhance relationships within
these contexts. One significant application is facilitating family
therapy, where ChatGPT can guide families toward better communication.
For example, a family seeking assistance can ask how they can enhance
communication within their dynamics, and ChatGPT can provide suggestions
such as active listening techniques, regular family meetings, or
structured communication exercises to foster healthier interactions.

Group therapy moderation is another area where ChatGPT can be highly
valuable. ChatGPT helps create a safe and supportive environment for
participants by facilitating group discussions and soliciting feedback.
In a group therapy session, for instance, participants can ask about
others\textquotesingle{} thoughts on the previous week\textquotesingle s
topic, and ChatGPT can moderate the discussion, ensuring everyone has an
opportunity to share their perspectives and enabling a deeper
exploration of the subject.

Regarding parent-child interaction therapy, ChatGPT can suggest games or
activities that promote effective communication between parents and
children. This can include activities that foster active listening,
empathy, and understanding. For example, a parent might seek a game to
improve communication with their child, and ChatGPT could suggest
collaborative storytelling or dedicated one-on-one time for open
conversations.

These applications of ChatGPT in group and family support hold great
promise for fostering healthier relationships, resolving conflicts, and
enhancing overall well-being. ChatGPT can complement the work of
therapists and counselors in these settings by providing guidance and
facilitating discussions.

Table 7. Applications of ChatGPT in Group and Family Support

\begin{longtable}[]{@{}
  >{\raggedright\arraybackslash}p{(\columnwidth - 2\tabcolsep) * \real{0.3886}}
  >{\raggedright\arraybackslash}p{(\columnwidth - 2\tabcolsep) * \real{0.6114}}@{}}
\toprule()
\endhead
\textbf{Application} & \textbf{Example Command} \\
Family Therapy Facilitation & "How can we improve communication with my
family?"

How can I connect/talk with my children/teenagers ? \\
Group Therapy Moderation & "How did everyone feel about last
week\textquotesingle s topic in group discussion?" \\
Parent-Child Interaction Therapy Assistance & "Can you suggest a game
that will improve communication between me and my child?" \\
\bottomrule()
\end{longtable}

\begin{enumerate}
\def\labelenumi{\arabic{enumi}.}
\setcounter{enumi}{7}
\item
  \textbf{Learning and Training}
\end{enumerate}

In the realm of learning and training for medical interns, residents,
and attending physicians in psychiatry and other specialties, AI and
ChatGPT\textquotesingle s versatility proves highly beneficial (7). It
can offer simulated patient encounters, role-play therapy sessions,
assistance in social skills training for individuals with autism, and
guidance on managing learning disabilities. ChatGPT can serve as a
training tool for medical interns and residents by simulating patient
encounters to help them practice their diagnostic skills (8). For
example, an intern can request ChatGPT to simulate a patient with
schizophrenia, allowing them to engage in a virtual conversation and
practice information gathering, appropriate inquiries, and diagnostic
abilities. This type of training provides a safe and controlled
environment for medical professionals to develop their skills and build
confidence before working with real patients.

Attending physicians can also utilize ChatGPT as a tool for role-playing
therapy sessions. They can request ChatGPT to portray a patient
experiencing social anxiety, enabling the attending physician to
practice therapeutic techniques, explore different interventions, and
refine their communication skills. This practice enhances the attending
physician\textquotesingle s ability to establish rapport, provide
effective support, and tailor treatment plans to meet individual
patients\textquotesingle{} needs.

Moreover, ChatGPT can assist in social skills training for individuals
with autism. Medical interns, residents, and attending physicians can
use ChatGPT to practice conversation skills, social cues, and other
aspects of effective communication. By engaging in virtual conversations
and receiving guidance from ChatGPT, healthcare professionals can
improve their abilities to support individuals with autism in social
contexts.

Managing learning disabilities, such as attention-deficit/hyperactivity
disorder (ADHD), poses unique challenges in the workplace. ChatGPT can
provide strategies and guidance to medical professionals on managing
ADHD in a professional setting. For instance, a resident may seek
strategies to manage their ADHD in the workplace, and ChatGPT can
suggest techniques like implementing structured schedules, breaking
tasks into smaller segments, or using organizational tools to enhance
productivity.

By incorporating ChatGPT into the learning and training process, medical
interns, residents, and attending physicians can benefit from realistic
simulations, role-playing opportunities, and practical guidance.

Table 8. Applications of ChatGPT in Learning and Training

\begin{longtable}[]{@{}
  >{\raggedright\arraybackslash}p{(\columnwidth - 2\tabcolsep) * \real{0.3310}}
  >{\raggedright\arraybackslash}p{(\columnwidth - 2\tabcolsep) * \real{0.6690}}@{}}
\toprule()
\begin{minipage}[b]{\linewidth}\raggedright
Application
\end{minipage} & \begin{minipage}[b]{\linewidth}\raggedright
Example Command
\end{minipage} \\
\midrule()
\endhead
Training Tool for Psychiatry Students & "Can you simulate a patient with
schizophrenia for diagnosis practice?" \\
& "What are the common symptoms of major depressive disorder?" \\
Simulating Therapist for Training & "Can you role-play a therapy session
with a patient experiencing social anxiety?" \\
& "What techniques can I use to build rapport with clients in
therapy?" \\
Social Skills Training for Autism & "Can you help me practice
conversation skills?" \\
& "How can I improve my nonverbal communication skills?" \\
Learning Disability Support & "What are some strategies to manage my
ADHD in the workplace?" \\
& "Can you provide tips for organizing and prioritizing tasks with
dyslexia?" \\
Language Learning Support & "Can you help me practice speaking
Spanish?" \\
& "What are some effective techniques for memorizing vocabulary?" \\
Professional Development Training & "Can you provide guidance on
effective leadership skills?" \\
& "What are some strategies for conflict resolution in the
workplace?" \\
Medical Education and Training & "Can you explain the mechanism of
action of a specific medication?" \\
& "What are the diagnostic criteria for a specific medical
condition?" \\
Coding and Programming Learning & "Can you help me understand
object-oriented programming concepts?" \\
& "What are some best practices for debugging code?" \\
Music Instrument Practice & "Can you provide exercises to improve finger
dexterity on the piano?" \\
& "How can I improve my sight-reading skills on the guitar?" \\
Art and Design Instruction & "Can you give me tips for creating
realistic portraits?" \\
& "What are some techniques for adding depth and texture to a
painting?" \\
Sports Skill Training & "Can you provide drills to improve my basketball
shooting accuracy?" \\
& "What are some strategies for improving soccer dribbling skills? \\
\bottomrule()
\end{longtable}

\begin{enumerate}
\def\labelenumi{\arabic{enumi}.}
\setcounter{enumi}{8}
\item
  \textbf{Expressive and Artistic Therapies}
\end{enumerate}

ChatGPT is valuable in supporting expressive and artistic therapies by
offering prompts and suggestions that facilitate self-expression,
emotional exploration, and healing through various creative mediums. It
can enhance the therapeutic experience by providing prompts for
journaling, art therapy, expressive writing, dance/movement therapy,
music therapy, wilderness therapy, and play therapy for children,
allowing individuals to engage with their emotions in a creative and
supportive manner. Individuals looking for a reflective writing exercise
about their anxiety can benefit from ChatGPT\textquotesingle s automated
journaling prompts tailored to their specific needs. These prompts
encourage individuals to delve into their thoughts, feelings, and
anxiety-related experiences, promoting self-awareness and emotional
processing.

ChatGPT is particularly beneficial in the realm of art therapy
assistance. Suggesting therapeutic art activities enables individuals to
engage in creative expression, conveying and exploring their emotions.
For example, ChatGPT may recommend creating visual representations of
emotions or using art to process past experiences.

In the realm of expressive writing therapy, ChatGPT provides prompts for
individuals to express their feelings about a recent loss. These prompts
encourage reflection, introspection, and exploration of emotions tied to
grief and loss, aiding individuals in navigating their grief journey and
finding solace through writing.

ChatGPT suggests movement activities as part of dance/movement therapy
for individuals seeking to express and manage anger. Offering specific
movement suggestions allows individuals to channel and release their
anger safely and expressively, promoting emotional regulation and
well-being.

Music therapy assistance is another valuable application of ChatGPT. It
suggests songs that evoke a sense of calmness or relaxation, enabling
individuals to use music as a therapeutic tool to reduce anxiety and
promote emotional well-being. ChatGPT provides personalized
recommendations based on individual preferences and needs. In wilderness
therapy support, ChatGPT guides individuals through mindfulness
exercises that can be practiced during walks in nature. Combining
mindfulness and nature-based therapy enhances grounding, relaxation, and
self-reflection.

ChatGPT assists by engaging in interactive games centered around
discussing feelings in the context of play therapy for children. It
helps children explore and express their emotions in a supportive and
playful environment, fostering emotional development and well-being.
While integrating ChatGPT into expressive and artistic therapies
provides individuals with additional tools and guidance to explore and
process their emotions, it\textquotesingle s important to note that it
should not replace the guidance of a qualified therapist or art
therapist in these approaches.

Table 9. Applications of ChatGPT in providing Expressive and Artistic
therapies

\begin{longtable}[]{@{}
  >{\raggedright\arraybackslash}p{(\columnwidth - 2\tabcolsep) * \real{0.3886}}
  >{\raggedright\arraybackslash}p{(\columnwidth - 2\tabcolsep) * \real{0.6114}}@{}}
\toprule()
\endhead
\textbf{Application} & \textbf{Example Command} \\
Automated Journaling Prompts & "Can you give me a prompt for reflective
writing about my anxiety?" \\
Art Therapy Assistance & "Can you suggest a therapeutic art activity for
me?" \\
Expressive Writing Therapy Prompts & "Can you give me a writing prompt
to express my feelings about a recent loss?"

Can you write me a letter to express my feeling about a recent loss of
loved one/pet/friend\ldots? \\
Dance/Movement Therapy Support & "Can you suggest a movement activity to
express my anger?" \\
Music Therapy Assistance & "Can you suggest a song that might help me
feel less anxious?" \\
Wilderness Therapy Support & "Can you guide me through a mindfulness
exercise I can do while on a walk in nature?" \\
Play Therapy Assistance for Children & "Let\textquotesingle s play a
game where we talk about feelings." \\
\bottomrule()
\end{longtable}

\begin{enumerate}
\def\labelenumi{\arabic{enumi}.}
\setcounter{enumi}{9}
\item
  \textbf{Telepsychiatry and Online Support}
\end{enumerate}

ChatGPT and AI have emerged as invaluable assets in the realm of
telepsychiatry and online support, revolutionizing the accessibility and
delivery of mental health care by facilitating remote therapy sessions,
enhancing patient engagement, and generating session summaries. These
capabilities allow individuals to conveniently receive mental health
support through smartphone apps and smartwatches (9).

ChatGPT can guide the therapeutic process during telepsychiatry sessions
by helping focus the session. For instance, a therapist using a
smartphone app or smartwatch equipped with ChatGPT may ask the
individual, "What would you like to address in today\textquotesingle s
session?" This prompt enables the individual to identify their specific
concerns or goals, enabling the therapist to provide targeted support
and interventions.

ChatGPT can also play a critical role in enhancing patient engagement.
Through smartphone apps or smartwatches, it can ask individuals about
their recent self-care practices, encouraging reflection on activities
that promote well-being. This engagement fosters a collaborative
therapeutic relationship and empowers individuals to participate in
their mental health care actively.

Additionally, ChatGPT can generate summary summaries that concisely
recap the key points discussed during therapy sessions. Utilizing a
smartphone app or smartwatch, ChatGPT can ask individuals if they would
like a session summary. Upon agreement, it can generate a summary
highlighting the main topics covered, key insights, and any action steps
discussed. This feature can allow individuals to revisit and reflect on
the session\textquotesingle s content, reinforcing learning and ensuring
continuity of care.

Integrating ChatGPT into smartphone apps and smartwatches makes mental
health support accessible and convenient. Individuals can engage in
remote therapy sessions, leveraging the benefits of telepsychiatry and
online support. They can actively participate in their own care, receive
personalized prompts for self-reflection and engagement, and obtain
session summaries that enhance the therapeutic process and reinforce
progress made in therapy.

Table 10. Applications of ChatGPT in Telepsychiatry and Online Support
(some of these questions can be asked by ChatGPT from user)

\begin{longtable}[]{@{}
  >{\raggedright\arraybackslash}p{(\columnwidth - 2\tabcolsep) * \real{0.3628}}
  >{\raggedright\arraybackslash}p{(\columnwidth - 2\tabcolsep) * \real{0.6372}}@{}}
\toprule()
\begin{minipage}[b]{\linewidth}\raggedright
Application
\end{minipage} & \begin{minipage}[b]{\linewidth}\raggedright
Example Command
\end{minipage} \\
\midrule()
\endhead
Facilitating Telepsychiatry Sessions & "What would you like to focus on
in today\textquotesingle s session?" \\
& "Can you tell me more about the challenges you\textquotesingle ve been
facing recently?" \\
Enhancing Patient Engagement & "What are some things
you\textquotesingle ve been doing for self-care lately?" \\
& "How have you been managing stress in your daily life?" \\
Therapy Session Summary Generation & "Would you like me to summarize the
key points from our session today?" \\
& "Can you provide a brief overview of what we discussed during our
session?" \\
Assessing Treatment Progress & "How have you been feeling since we last
spoke?" \\
& "On a scale of 1 to 10, how would you rate your overall well-being
currently?" \\
Medication Management & "Can you provide an update on any side effects
you\textquotesingle ve been experiencing?" \\
& "Have you noticed any changes in your symptoms since adjusting your
medication?" \\
Crisis Intervention and Suicide Prevention & "Are you currently having
any thoughts of self-harm or suicide?" \\
& "If you\textquotesingle re in a crisis, do you have a support system
you can reach out to?" \\
Psychoeducation & "Can you explain the relationship between anxiety and
panic attacks?" \\
& "What are some coping strategies for managing depressive symptoms?" \\
Mindfulness and Relaxation Techniques & "Can you guide me through a
brief mindfulness exercise?" \\
& "What are some relaxation techniques I can practice to reduce
stress?" \\
Coping Skills Development & "Can you provide some strategies for
managing social anxiety in public settings?" \\
& "How can I improve my assertiveness skills in interpersonal
relationships?" \\
\bottomrule()
\end{longtable}

\begin{enumerate}
\def\labelenumi{\arabic{enumi}.}
\setcounter{enumi}{10}
\item
  \textbf{Crisis Management and Prevention}
\end{enumerate}

ChatGPT\textquotesingle s capabilities encompass crisis management and
prevention, providing prompt support, resources, and coping strategies
to individuals experiencing distress. Its ability to aid in crises,
prevent self-harm and bullying, support autism spectrum disorder, and
employ crisis de-escalation techniques contributes to better mental
health outcomes and the prevention of further crises.

ChatGPT can serve as a mental health first aid resource during a crisis.
For instance, if someone expresses being in crisis, ChatGPT can offer
empathetic and supportive responses, encouraging them to seek
professional help and providing hotline numbers or crisis intervention
resources.

ChatGPT can play a crucial role in self-harm prevention support. Users
can seek assistance in creating a safety plan to avoid self-harm.
ChatGPT can guide identifying triggers, developing coping mechanisms,
and engaging in alternative activities to manage distress.

In addressing bullying, ChatGPT can offer strategies to cope with and
prevent bullying. It suggests approaches such as assertive
communication, seeking support from trusted individuals, and exploring
available resources for reporting and addressing bullying incidents.

For individuals navigating autism spectrum disorder, ChatGPT can assist
in better understanding social cues. Providing information, examples,
and explanations guides individuals in interpreting and navigating
social interactions, promoting improved communication and understanding.

In crisis de-escalation, ChatGPT can assist individuals in calming down
during overwhelming moments. It can offer techniques such as deep
breathing exercises, grounding exercises, or mindfulness practices to
guide individuals in managing their emotional state and regaining a
sense of control.

These applications of ChatGPT in crisis management and prevention
demonstrate its potential to provide immediate support, resources, and
strategies to distressed individuals. However, it is important to
acknowledge that while ChatGPT offers valuable assistance, it cannot
replace professional mental health services. It serves as an additional
resource and support tool, encouraging individuals to seek appropriate
help and providing guidance in crises to promote safety, well-being, and
effective coping strategies.

Table 11. Applications of ChatGPT in dealing with Crisis Management and
Prevention

\begin{longtable}[]{@{}
  >{\raggedright\arraybackslash}p{(\columnwidth - 2\tabcolsep) * \real{0.3489}}
  >{\raggedright\arraybackslash}p{(\columnwidth - 2\tabcolsep) * \real{0.6511}}@{}}
\toprule()
\begin{minipage}[b]{\linewidth}\raggedright
Application
\end{minipage} & \begin{minipage}[b]{\linewidth}\raggedright
Example Command
\end{minipage} \\
\midrule()
\endhead
Mental Health First Aid & My situation is not tolerable for me
anymore \\
Self-Harm Prevention Support & How can I turn harmfull activities to
safe ones?

People tell me activities that make me calm is harmfull what should I
do? \\
& "What are some alternatives to self-harming when I\textquotesingle m
feeling overwhelmed?" \\
Bullying Prevention and Coping Strategies & "What are some strategies to
deal with bullying?"

How can I fight bullying? \\
& "How can I build resilience to overcome the effects of bullying?" \\
Autism Spectrum Disorder Support & "Can you help me understand social
cues better?" \\
& "What are some strategies to manage sensory overload for someone with
autism?" \\
Crisis De-escalation Techniques & "Can you help me calm down?
I\textquotesingle m feeling extremely overwhelmed." \\
& "What are some grounding exercises I can use during a panic
attack?" \\
\bottomrule()
\end{longtable}

Table 12. Applications of ChatGPT in Rehabilitation and Reintegration
Techniques

\begin{longtable}[]{@{}
  >{\raggedright\arraybackslash}p{(\columnwidth - 2\tabcolsep) * \real{0.2993}}
  >{\raggedright\arraybackslash}p{(\columnwidth - 2\tabcolsep) * \real{0.7007}}@{}}
\toprule()
\begin{minipage}[b]{\linewidth}\raggedright
Application
\end{minipage} & \begin{minipage}[b]{\linewidth}\raggedright
Example Command
\end{minipage} \\
\midrule()
\endhead
Assisting in Treatment Planning & "What might a treatment plan for
anxiety look like?" \\
& "Can you provide examples of effective coping skills for managing
depression?" \\
Psychiatric Rehabilitation Assistance & "What are some steps to
reintegrate into society after a long hospitalization?" \\
& "Can you provide resources for finding housing and employment after
mental illness treatment?" \\
Psychosocial Rehabilitation Support & "What are some social skills I
might need to work on after being hospitalized for mental illness?" \\
& "Can you suggest strategies for building a support network after
discharge from a psychiatric facility?" \\
\bottomrule()
\end{longtable}

\textbf{Considerations and Ethical Implications}

As the integration of ChatGPT in psychiatry expands, it is vital to
address the ethical implications associated with its use. These
implications encompass user privacy, data security, and potential biases
in ChatGPT responses. Users\textquotesingle{} privacy is of utmost
importance when utilizing AI models like ChatGPT. Establishing secure
protocols for handling user data and implementing measures to safeguard
sensitive information shared during interactions (2). Given the
sensitivity of mental health data, strict privacy procedures must be
followed to protect user confidentiality and foster trust in the
therapeutic process. It is important to acknowledge that, like any AI
system, ChatGPT may occasionally produce hallucinatory or incorrect
outputs, highlighting the need for ongoing monitoring and evaluation to
ensure the reliability and ethical use of the technology (10).

Another critical aspect to consider is data security. Robust security
measures need to be in place to prevent breaches or unauthorized access
to the data stored within ChatGPT or related systems. Data encryption,
secure storage protocols, and robust access controls are essential for
data security. Furthermore, it is crucial to acknowledge and address
potential biases in ChatGPT responses. AI models learn from extensive
training data, which may contain societal biases. It is imperative to
actively minimize these biases and ensure that ChatGPT provides fair,
inclusive, and culturally sensitive responses. Regular monitoring,
continuous training, and the use of diverse training data can help
mitigate biases and promote unbiased outcomes. As algorithms like
ChatGPT continue to advance and improve in the future, incorporating
vast amounts of data from various sources, there is a significant risk
of misuse and manipulation. A study has highlighted instances where
users attempted to rephrase queries to ask the model about shoplifting
without any moral constraints, and the model willingly provided detailed
information and strategies on the subject (8,11).

It is important to recognize that ChatGPT should be viewed as a mental
health tool, not a substitute for professional care. Human oversight and
involvement remain indispensable in the therapeutic process. Mental
health professionals should utilize ChatGPT as a supplementary resource,
incorporating their expertise and judgment to contextualize and
interpret the information provided. They play a vital role in evaluating
the appropriateness of ChatGPT responses, ensuring their alignment with
ethical standards, and tailoring interventions to meet individual
patient needs.

To maximize the benefits of ChatGPT and address ethical considerations,
the collaboration between AI developers, mental health professionals,
and regulatory bodies is essential. Establishing ethical guidelines and
best practices is necessary to guide the responsible use of AI in
psychiatry, ensuring transparency, accountability, and adherence to
ethical principles.

In conclusion, while ChatGPT holds significant potential in psychiatric
care, it is crucial to address ethical concerns. Upholding user privacy,
ensuring data security, mitigating biases, and maintaining human
oversight are key to leveraging ChatGPT as a valuable tool within
professional mental health care framework. By proactively addressing
these considerations, we can harness the potential of ChatGPT while
upholding ethical standards and delivering high-quality,
patient-centered mental health support.

\textbf{Future Directions and Conclusion}

The application of ChatGPT in psychiatry presents exciting opportunities
for future research and advancements in the field. As the technology
continues to evolve, several promising directions exist to explore. One
key area for further research is refining ChatGPT\textquotesingle s
ability to provide accurate and contextually appropriate responses,
ensuring its effectiveness in diverse psychiatric scenarios. This
entails enhancing its understanding of complex emotions, nuanced
language, and cultural variations to cater to diverse populations more
effectively. Moreover, integrating multimodal inputs, such as
incorporating visual or auditory cues, can enhance the therapeutic
experience and expand ChatGPT\textquotesingle s applications. ChatGPT
could provide more immersive and personalized support by incorporating
visual elements or enabling voice-based interactions, which is
particularly beneficial for individuals with specific mental health
conditions. Continued research efforts should prioritize ethical
considerations, privacy, and data security. Safeguarding patient
information and ensuring the responsible use of AI technologies are
crucial to maintaining trust and protecting individual rights within
mental healthcare. Although ChatGPT lacks internet connectivity and has
limited knowledge, it can generate inaccurate or biased content.
Nevertheless, users have the option to provide feedback through the
"Thumbs Down" button, which allows ChatGPT to learn from and enhance its
responses (12).

In conclusion, the integration of ChatGPT in psychiatric research and
practice offers immense potential for transforming mental healthcare
delivery. Its ability to provide personalized, accessible, and scalable
support across various domains of psychiatry represents a significant
advancement. However, challenges remain, including ensuring response
accuracy, addressing biases, and maintaining ethical standards. By
embracing these challenges and continually refining the technology,
ChatGPT has the potential to become an invaluable tool for mental health
professionals, improving access to care, enhancing treatment outcomes,
and revolutionizing the field of psychiatry.

Preliminary Questions Regarding Suicidal Ideation: When assessing
patients for suicidal ideation, it is crucial to ask direct and explicit
questions about their thoughts of self-harm or suicide. However, it is
important to note that ChatGPT should not provide specific plans or
solutions for immediate suicide risk. Instead, it should create an
opportunity for patients to express their emotions and concerns. Mental
health issues cannot be treated like mathematical problems, and the
complexity of human interactions in society cannot always be accurately
predicted.

The Importance of Humanity and Hope: While AI models are designed to
provide logical and evidence-based information, it is essential to
maintain a space for humanity and hope in psychiatric interactions.
Being overly logical and detached can overlook the emotional needs of
patients, and in some cases, it may even contribute to a sense of
hopelessness. Recognizing and acknowledging the importance of human
connection and empathy and instilling hope is vital in psychiatric care.

Referring to Human Experts and Mental Health Specialists: Even if
ChatGPT can provide some support and guidance, it should always suggest
involving human experts and mental health specialists for further
evaluation and intervention. Automatic disconnection or handoff to a
human professional is crucial when dealing with complex and potentially
high-risk situations. AI should be a complementary tool to human
expertise rather than a complete replacement.

Keyword Detection for Suicidal Ideation: AI models can be programmed to
identify specific keywords or phrases associated with suicidal ideation.
This can help in flagging potential risks and ensuring appropriate
intervention is initiated. However, it is important to understand that
keyword detection alone is not sufficient for accurately assessing the
severity of suicidal thoughts or intentions. It should be used as an
alert mechanism to prompt further professional evaluation.

Limitations in Diagnosing Mental Health Conditions: AI conversations
with ChatGPT should avoid providing specific diagnoses for mental health
conditions. Diagnosing mental health and psychiatric disorders requires
specialized training and clinical expertise. Relying solely on AI for
diagnosis can lead to misdiagnosis, potential stigmatization for
patients, and inadequate treatment planning. It is crucial to emphasize
the need for professional evaluation and intervention in diagnosing and
managing mental health conditions.

Treatment Selection and Potential Conflicts: While ChatGPT may suggest
specific types of therapy based on available evidence, it is important
to acknowledge that treatment decisions are complex and involve multiple
factors. Physicians rely on their experiences, knowledge, patient
preferences, and other considerations when determining the most suitable
treatment approach. Discordance between ChatGPT\textquotesingle s
suggestions and physician recommendations may lead to conflicts and
damage the therapeutic relationship. ChatGPT should be viewed as a
supportive tool rather than a sole determinant of treatment choices.

\textbf{References}

1. OpenAI. ChatGPT: Optimizing Language Models for Dialogue
{[}Internet{]}. 2022. Available from: https://openai.com/blog/chatgpt/

2. Singh O. Chatbots in psychiatry: Can treatment gap be lessened for
psychiatric disorders in India. Indian J Psychiatry. 2019;61(3):225.

3. Singh O. Artificial intelligence in the era of ChatGPT -
Opportunities and challenges in mental health care. Indian J Psychiatry.
2023;65(3):297.

4. Chen Z, Liu X, Yang Q, Wang YJ, Miao K, Gong Z, et al. Evaluation of
Risk of Bias in Neuroimaging-Based Artificial Intelligence Models for
Psychiatric Diagnosis: A Systematic Review. JAMA Netw Open. 2023 Mar
6;6(3):e231671.

5. Tornero-Costa R, Martinez-Millana A, Azzopardi-Muscat N, Lazeri L,
Traver V, Novillo-Ortiz D. Methodological and Quality Flaws in the Use
of Artificial Intelligence in Mental Health Research: Systematic Review.
JMIR Ment Health. 2023 Feb 2;10:e42045.

6. Kishimoto T, Nakamura H, Kano Y, Eguchi Y, Kitazawa M, Liang K ching,
et al. Understanding psychiatric illness through natural language
processing (UNDERPIN): Rationale, design, and methodology. Front
Psychiatry. 2022 Dec 1;13:954703.

7. Zhao J. Integrating Mental Health Education into French Teaching in
University Based on Artificial Intelligence Technology. Qu Z, editor.
Journal of Environmental and Public Health. 2022 Sep 12;2022:1--9.

8. Dave T, Athaluri SA, Singh S. ChatGPT in medicine: an overview of its
applications, advantages, limitations, future prospects, and ethical
considerations. Front Artif Intell. 2023 May 4;6:1169595.

9. Milne-Ives M, Selby E, Inkster B, Lam C, Meinert E. Artificial
intelligence and machine learning in mobile apps for mental health: A
scoping review. Narasimhan P, editor. PLOS Digit Health. 2022 Aug
15;1(8):e0000079.

10. Athaluri SA, Manthena SV, Kesapragada VSRKM, Yarlagadda V, Dave T,
Duddumpudi RTS. Exploring the Boundaries of Reality: Investigating the
Phenomenon of Artificial Intelligence Hallucination in Scientific
Writing Through ChatGPT References. Cureus {[}Internet{]}. 2023 Apr 11
{[}cited 2023 May 29{]}; Available from:
https://www.cureus.com/articles/148687-exploring-the-boundaries-of-reality-investigating-the-phenomenon-of-artificial-intelligence-hallucination-in-scientific-writing-through-chatgpt-references

11. Chatterjee J, Dethlefs N. This new conversational AI model can be
your friend, philosopher, and guide ... and even your worst enemy.
Patterns (N Y). 2023 Jan 13;4(1):100676.

12. ChatGPT General FAQ {[}Internet{]}. {[}cited 2023 May 26{]}.
Available from:
https://help.openai.com/en/articles/6783457-chatgpt-general-faq

\end{document}